\documentclass[12pt]{article}
\usepackage{amsmath,amssymb}
\usepackage{hyperref}
\usepackage{units}
\usepackage{color}
\usepackage[T1]{fontenc}
\usepackage[utf8]{inputenc}
\usepackage{authblk}
\usepackage{bm}
\usepackage{graphicx}
\title{On approximating the free harmonic oscillator by a particle in a box
\thanks{%
PACS:03.65.Ge} 
}
\author[]{Kunle Adegoke\thanks{Corresponding author: adegoke00@gmail.com}}
\author[]{Adenike Olatinwo}
\author[]{Henry Otobrise}
\author[]{\mbox{Rauf Giwa}}
\author[]{Gbenga Olunloyo}
\affil{Department of Physics and Engineering Physics, \mbox{Obafemi Awolowo University}, Ile-Ife, Nigeria}

\begin{document}

\date{}

\maketitle

\begin{abstract}
\noindent The main purpose of this paper is to demonstrate and illustrate, once again, the potency of the variational technique as an approximation procedure for the quantization of quantum mechanical systems. By choosing particle-in-a-box wavefunctions as trial wavefunctions, with the size of the box as the variation parameter, approximate eigenenergies and the corresponding eigenfunctions are obtained for the one dimensional free harmonic oscillator.
\end{abstract}

\tableofcontents
\listoffigures
\section{Introduction}
This paper was inspired by the 1965 work of Padnos, as reported in reference~\cite{padnos}. Using the variation ansatz with normalized wavefunctions of a particle in a box as trial wavefunctions, he obtained approximate values for the ground state energy and the first excited state energy of the one dimensional quantum harmonic oscillator. Padnos' work was preceded by and is an improvement upon the work of Rich~\cite{rich}, who approximated the oscillator by a particle in a one dimensional box whose size was the classical range of the oscillator. 

\bigskip

One of the reasons Padnos was discouraged from extending the calculations to higher energy levels was the belief that the calculation becomes more tedious as one must introduce extra terms to orthogonalize the new function to the ones already found (for example, the second excited state must be chosen to be orthogonal to both the ground state and the first excited state). The authors of this present paper have no need to introduce any extra terms since the eigenstates of the particle in a box can all be chosen to be mutually orthogonal, as simple trigonometric functions (see section~\ref{sec.pf1lpk4}).

\bigskip

Another notable paper which employed the variational method for harmonic oscillator quantization is reference~\cite{casaubon}, where different sets of basis functions, built from non-orthogonal monomials, were used as trial wavefunctions.

\bigskip

A particle of mass $m$, free to move {\em only} in a `box' of size $2L$, so that the potential, $V(x)$, is
\[
V(x)=
\begin{cases}
0\,,\quad & |x|\le L\\
\\
\infty\,, & \quad |x|> L
\end{cases}\,,
\]
is decribed by the  Hamiltonian
\[
H_F=T(x)=-\frac{\hbar^2}{2m}\frac{\partial^2}{\partial {x^2}},\quad |x|\le L\,.
\]
The eigenfunctions of $H_F$ are non-degenerate and are given by
\[
\phi_r(x)=
\begin{cases}
\sqrt{\frac 1L}\cos{\left[\frac{(r+1)\pi x}{2L}\right]}, & r=0,2,4,\ldots\\
\\
\sqrt{\frac 1L}\sin{\left[\frac{(r+1)\pi x}{2L}\right]}, & r=1,3,5,\ldots\,,
\end{cases}
\]
with corresponding eigenvalues
\begin{equation}\label{equ.npvmv9n}
\varepsilon_r=(r+1)^2\varepsilon\,,
\end{equation}
where $\varepsilon=\varepsilon_0=\hbar^2\pi^2/8mL^2$.
The eigenfunctions $\phi_r(x)$ can be written more compactly as
\begin{equation}\label{equ.hunt4ni}
\phi _r (x ) = \sqrt {\frac{1}{L}} \cos \left[ {\frac{\pi }{2}\sin^2 \left(\frac{{r\pi }}{2}\right) - \frac{{(r+1)\pi x }}{2L}} \right],\quad r=0,1,2,3,\ldots
\end{equation}
If the particle is not free in the box but is instead under the influence of a potential, $V(x)$, where
\[
V(x)=
\begin{cases}
\frac12m\omega^2x^2\,, &\quad |x|\le L\\
\\
\infty\,, & \quad |x|> L
\end{cases}\,,
\]
then the system becomes a `confined' quantum harmonic oscillator (CHO), and is now described by the Hamiltonian
\[
H_{CHO}:=-\frac{\hbar^2}{2m}\frac{\partial^2}{\partial {x^2}}+\frac12m\omega^2x^2,\quad |x|\le L\,,
\]
where $\omega$ is the classical frequency of the oscillator. Although the confined quantum harmonic oscillator has been studied for a long time (see references~\cite{baijal, campoy, gueorguiev, jaber, montgomery, marsiglio} and the references therein), it does not seem to enjoy the same popularity as its $L\to\infty$ limit, the `free' quantum harmonic oscillator (FHO), with
\[
H_{FHO}:=-\frac{\hbar^2}{2m}\frac{\partial^2}{\partial {x^2}}+\frac12m\omega^2x^2,\quad -\infty<x<\infty\,.
\]
It is our aim in this paper to apply the variation procedure to quantize $H_{FHO}$ through $H_{CHO}$. Since $H_F$ and $H_{CHO}$ live in the same Hilbert space we will use the complete set of functions $\phi_r(x)$ (eigenstates of $H_F$) as trial wavefunctions in the variation ansatz. The size of the box, $L$, then enters, naturally, as the variation parameter.

\section{The free harmonic oscillator as the variational limit of the confined harmonic oscillator}

\subsection{The choice of trial wavefunctions}\label{sec.pf1lpk4}
The functions $\phi_r(x)$ as given in~\eqref{equ.hunt4ni}, being non-degenerate eigenstates of a Hermitian operator, $H_F$, are a complete set of vectors spanning an infinite dimensional Hilbert space. In particular $\phi_r(x)$ being mutually orthogonal in the interval \mbox{$|x|\le L$} and satisfying the boundary conditions \mbox{$\phi_r(\pm L)=0$} are a suitable choice of trial wavefunctions for the confined harmonic oscillator Hamiltonian, $H_{CHO}$, whose eigenfunctions are also required to satisfy the same boundary conditions. The box size, $L$, then serves as the variation parameter which will be optimized to get the approximate eigenfunctions and eigenvalues of the free harmonic oscillator. Henceforth, any quantity derived for the free harmonic oscillator by optimization will be indicated with an asterik on its symbol. 

\bigskip

The optimized box size, ${L^*}$, is obtained by solving
\[
\left. { {\frac{\partial E_r}{{\partial L}} } } \right|_{L = {L^*}}  = 0
\]
for ${L^*}$, where
\begin{equation}\label{equ.j9s8h2w}
E_r=\left\langle {\phi _r } \right|H_{CHO}\left| {\phi _r } \right\rangle=\int_{ - L}^L {\phi _r H_{CHO}\phi _r\,dx}\,.
\end{equation}
The optimized approximate energy eigenvalues of the free harmonic oscillator are then given by
\begin{equation}\label{equ.k2hq971}
{E_r}^*  = \left. {E_r } \right|_{L = {L^*}}\,. 
\end{equation}
Actually, the $\varepsilon$ introduced in~\eqref{equ.npvmv9n} (ground state energy of the free particle in a box) is more convenient to use as the variation parameter, with its optimum value $\varepsilon^*$ being related to the optimum value of the box size ${L^*}$ by 
\begin{equation}\label{equ.w1gdgnk}
{L^*}{}^2=\frac{\hbar^2\pi^2}{8m\varepsilon^*}\,.
\end{equation}
$\varepsilon^*$ is obtained by expressing the $E_r$ of~\eqref{equ.j9s8h2w} in terms of $\varepsilon$ and then solving
\begin{equation}\label{equ.lovmei7}
\left. { {\frac{\partial E_r}{{\partial\varepsilon}} } } \right|_{\varepsilon = \varepsilon^*}  = 0
\end{equation}
for $\varepsilon^*$. The optimized approximate eigenenergies of the free harmonic oscillator are then obtained from
\begin{equation}\label{equ.k2hq971}
{E_r}^*  = \left. {E_r } \right|_{\varepsilon = \varepsilon^*}\,. 
\end{equation}
\subsection{Approximate eigenfunctions and eigenvalues for the free harmonic oscillator}
Performing the integration in~\eqref{equ.j9s8h2w} we have
\begin{equation}\label{equ.ux2llvh}
E_r  = (r + 1)^2 \varepsilon  + \frac{1}{\varepsilon }\frac{{\varepsilon _\omega{}^2 }}{8}\left( {\frac{{\pi ^2 }}{6} - \frac{1}{{(r + 1)^2 }}} \right)\,,\quad r=0,1,2,\ldots\,,
\end{equation}
where $\varepsilon_\omega=\hbar\omega$. Thus
\begin{equation}\label{equ.ycxzxep}
E_r=A_r\varepsilon+\frac{B_r}{\varepsilon}\,,
\end{equation}
where
\[
A_r  = (r + 1)^2\,,\quad B_r  = \frac{{\varepsilon _\omega ^2 }}{8}\left( {\frac{{\pi ^2 }}{6} - \frac{1}{{(r + 1)^2 }}} \right)\,.
\]
From~\eqref{equ.ux2llvh} and~\eqref{equ.lovmei7} we have
\[
{\varepsilon^*{}}^2=\frac{B_r}{A_r}\,,
\]
so that
\begin{equation}\label{equ.qvb8wqf}
\frac{\varepsilon^*}{\varepsilon _\omega}  =   \sqrt { {\frac{{\pi ^2 (r + 1)^2  - 6}}{{48(r + 1)^4 }}} }=\gamma(r)\,. 
\end{equation}
From~\eqref{equ.k2hq971} and~\eqref{equ.ycxzxep} we have
\[
{E_r}^*  = 2\sqrt {A_r B_r }\,,
\]
so that the approximate eigenenergies of the free harmonic oscillator, ${E_r}^*$, are given by
\begin{equation}\label{equ.jql5eex}
\boxed{{E_r}^*  = \varepsilon _\omega  \sqrt {\frac{{\pi ^2 (r + 1)^2  - 6}}{{12}}}\,,\quad r=0,1,2,3,\ldots.}
\end{equation}
Writing~\eqref{equ.w1gdgnk} as 
\[
\frac{1}{{{L^*}^2 }} = \frac{{8\alpha }}{{\pi ^2 }}\frac{{\varepsilon^* }}{{\varepsilon _\omega  }}\,,
\]
where $\alpha=m\omega/\hbar$, and using~\eqref{equ.qvb8wqf} we have
\begin{equation}\label{equ.gnp1cqh}
\frac{1}{{{L^*}}(r)} = \sqrt{\left( {\frac{{8\alpha }}{{\pi ^2 }}\gamma (r)} \right)}\,,
\end{equation}
so that the approximate eigenstates of the free harmonic oscillator are given by 
\begin{equation}\label{equ.g90qo75}
\boxed{{\phi_r}^* (x ) = \sqrt {\frac{1}{{L^*}(r)}} \cos \left[ {\frac{\pi }{2}\sin^2 \left(\frac{{r\pi }}{2}\right) - \frac{{(r+1)\pi x }}{2{L^*}(r)}} \right],\quad r=0,1,2,3,\ldots}
\end{equation}
The optimized eigenfunctions ${\phi_r}^* (x )$ are expected to approximate the eigenstates of the free harmonic oscillator in the interval: \mbox{$-L^*(r)\le x\le L^*(r)$}.
\subsection{Comparison with the exact results}
The free harmonic oscillator is one of the few quantum mechanical systems that can be quantized exactly. Expressions for the eigenenergies and the corresponding wavefunctions are derived in every book on quantum mechanics. For quantum numbers $r=0,1,2,\ldots$, the energy eigenvalues are given by
\begin{equation}\label{equ.w3990g8}
E_r=\left(r+\frac{1}{2}\right)\varepsilon_\omega\,,
\end{equation}
with corresponding normalized wavefunctions
\begin{equation}\label{equ.nb49ait}
\phi _r (x) = \left( {\frac{\alpha }{\pi }} \right)^{1/4} \frac{1}{{\sqrt {2^r r!} }}H_r (x\sqrt \alpha  )\exp ( - \alpha x^2 /2)\,,
\end{equation}
where $H_q(y)$ is the degree $q$ Hermite polynomial in $y$ and, as before, \mbox{$\alpha=m\omega/\hbar$} and \mbox{$\varepsilon_\omega=\hbar\omega$}. The first few Hermite polynomials are:
\[
\begin{split}
&H_0(y)=1,\quad H_1(y)=2y\,,\\
&H_2(y)=4y^2-2,\quad H_3(y)=8y^3-12y\,.
\end{split}
\]
\subsubsection{Comparison of energy eigenvalues}
\begin{figure}[h!]
\centering
\includegraphics[scale=.8,clip=true]{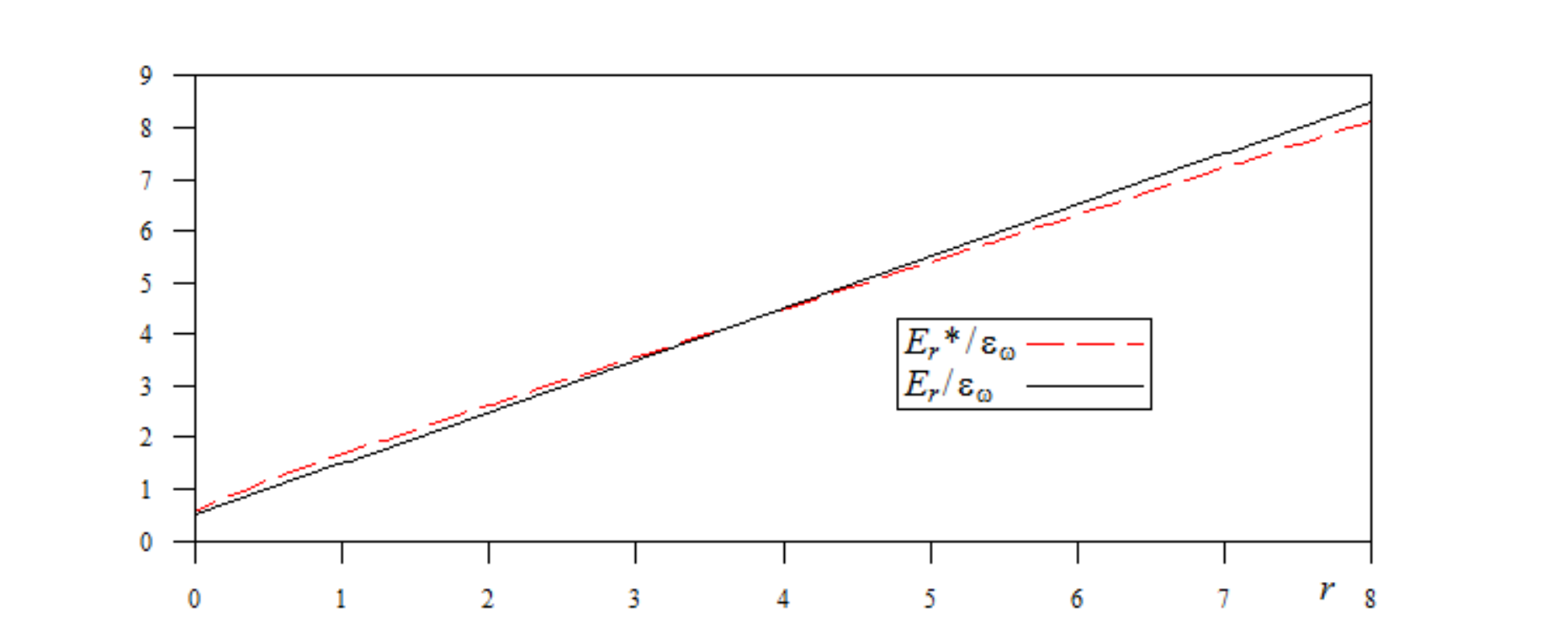}
\caption{Comparison of the approximate eigenenergies, $E_r^*$, with the exact eigenenergies, $E_r$, for the free harmonic oscillator.}
\label{fig.zsxuwpe}
\end{figure}

We recall expressions~\eqref{equ.jql5eex} and~\eqref{equ.w3990g8} for the approximate and exact eigenenergies of the free one-dimensional linear harmonic oscillator:
\[
\begin{split}
{E_r}^*  &= \varepsilon _\omega  \sqrt {\frac{{\pi ^2 (r + 1)^2  - 6}}{{12}}}\,,\\
&\qquad\qquad\qquad\qquad\qquad\qquad r=0,1,2,\ldots\\
E_r  &= \varepsilon _\omega  \left( {r + \frac{1}{2}} \right)\,.
\end{split}
\]
${E_r}^*/\varepsilon_\omega$ and $E_r/\varepsilon_\omega$ are plotted in Figure~\ref{fig.zsxuwpe} as functions of the harmonic oscillator quantum number~$r$. The agreement between ${E_r}^*$ and $E_r$ is quite remarkable, especially for low quantum numbers. 
\pagebreak
\subsubsection{Comparison of eigenfunctions}
To illustrate the agreement between the variation results and the exact wavefunctions for the free harmonic oscillator, it is instructive to plot the approximate and the exact wavefunctions together, as functions of position. Here we do this for the ground state and the first excited state.

\bigskip

\begin{figure}[h!]
\centering
\includegraphics[scale=.8,clip=true]{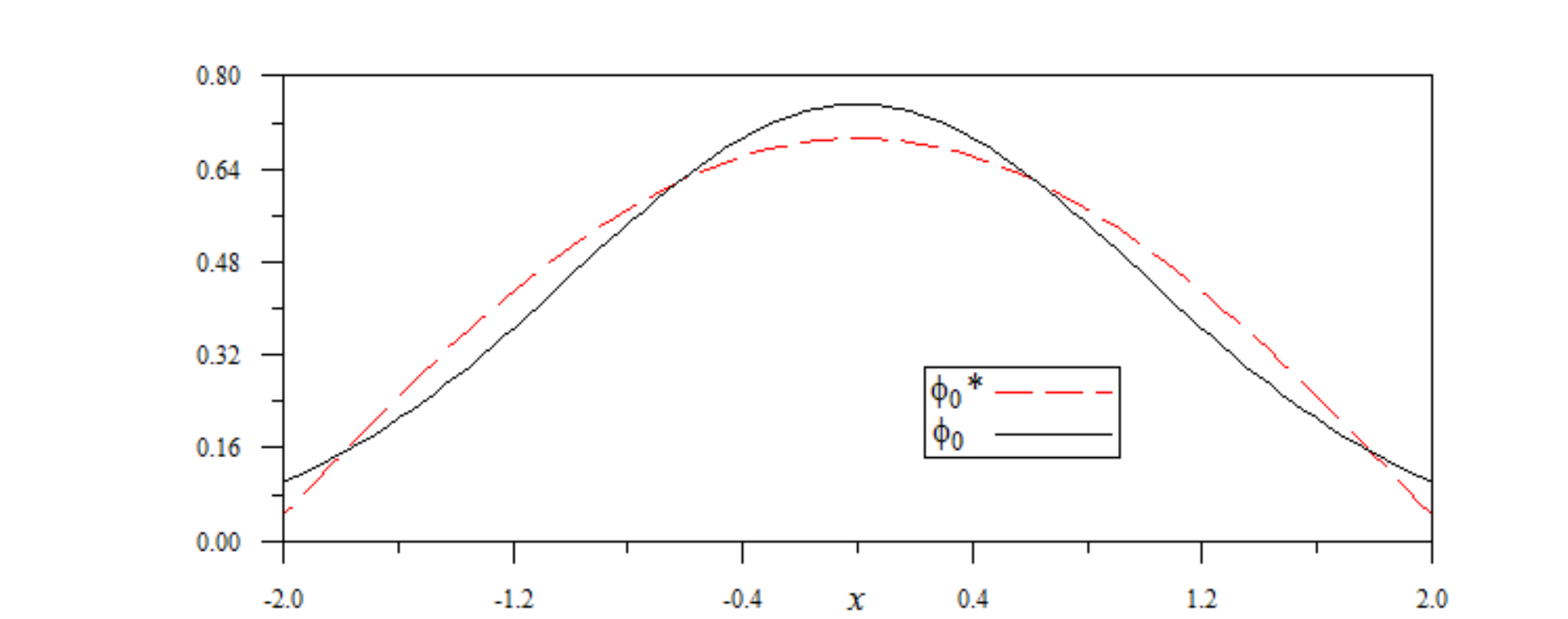}
\caption{Approximate ground state, $\phi_0^*(x)$, and exact ground state, $\phi_0(x)$ as functions of position, for the free harmonic oscillator, for $\alpha=1$.}
\label{fig.q96gfpe}
\end{figure}

From~\eqref{equ.g90qo75} and~\eqref{equ.nb49ait} we have
\[
{\phi_0}^* (x) = \left( {\frac{\alpha }{\pi }} \right)^{1/4} \left( {\frac{2}{\pi }} \right)^{1/4} \left( {\frac{{\pi ^2  - 6}}{3}} \right)^{1/8} \cos \left\{ {\left( {\frac{\alpha }{2}} \right)^{1/2} \left( {\frac{{\pi ^2  - 6}}{3}} \right)^{1/4} x} \right\}
\]
and
\[
\phi _0 (x) = \left( {\frac{\alpha }{\pi }} \right)^{1/4} \exp \left( { - \frac{{\alpha x^2 }}{2}} \right)\,.
\]
The good correlation between ${\phi_0}^* (x)$ and $\phi _0 (x)$ is already obvious from the Taylor series expansion of both functions:
\[
{\phi_0}^* (x) = \left( {\frac{\alpha }{\pi }} \right)^{1/4} \left( {\frac{2}{\pi }} \right)^{1/4} \left( {\frac{{\pi ^2  - 6}}{3}} \right)^{1/8}  + O(x^2 )
\]
and
\[
\phi _0 (x) = \left( {\frac{\alpha }{\pi }} \right)^{1/4}  + O(x^2 )\,,
\]
with
\[
\frac{{{\phi_0}^* (0)}}{{\phi _0 (0)}} = \left( {\frac{2}{\pi }} \right)^{1/4} \left( {\frac{{\pi ^2  - 6}}{3}} \right)^{1/8}  = 0.9221215996
\]
The approximate ground state wavefunction ${\phi_0}^*(x)$ and the exact ground state wavefunction $\phi_0(x)$ of the free harmonic oscillator, with $\alpha=1$ are shown in Figure~\ref{fig.q96gfpe} as functions of position.

\bigskip

As for the first excited state, we have from~\eqref{equ.g90qo75} and~\eqref{equ.nb49ait} we have
\[
\phi _1^ *  (x) = \left( {\frac{\alpha }{\pi }} \right)^{1/4} \left( {\frac{{2\pi ^2  - 3}}{{6\pi ^2 }}} \right)^{1/8} \sin \left\{ {\alpha ^{1/2} \left( {\frac{{2\pi ^2  - 3}}{6}} \right)^{1/4} x} \right\}
\]
and
\[
\phi _1 (x) = \left( {\frac{\alpha }{\pi }} \right)^{1/4} x\sqrt {2\alpha } \exp \left( { - \frac{{\alpha x^2 }}{2}} \right)\,.
\]
\begin{figure}[h!]
\centering
\includegraphics[scale=.8,clip=true]{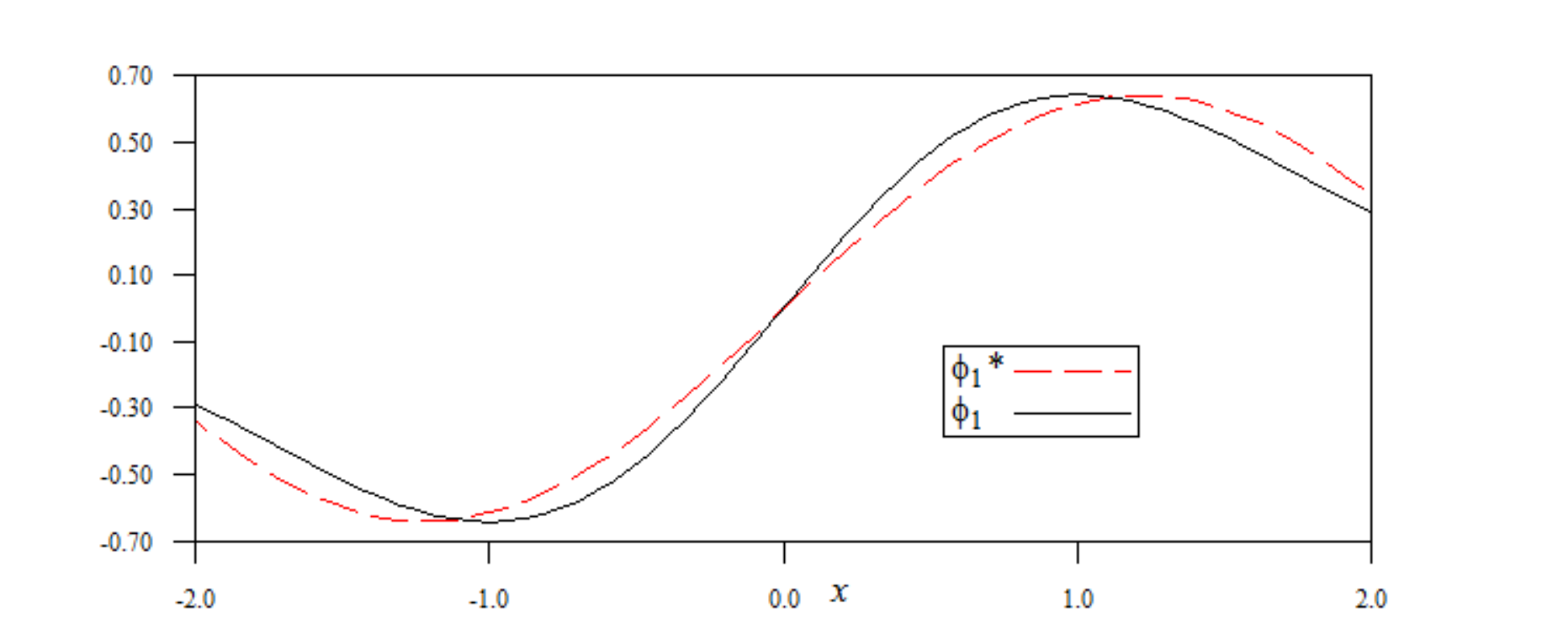}
\caption{Approximate first excited state, $\phi_1^*(x)$ and exact first excited state, $\phi_1(x)$, as functions of position, for the free harmonic oscillator, for $\alpha=1$.}
\label{fig.wipsqc4}
\end{figure}

The variation of $\phi _1^ *  (x)$ and $\phi _1 (x)$ with respect to position are as shown in Figure~\ref{fig.wipsqc4}.
\section{Summary and conclusion}
By using the normalized mutually orthogonal wavefunctions of the free-particle-in-a-box model as trial wavefunctions in variation calculation we have obtained approximate energy eigenvalues and the corresponding eigenstates for the one dimensional free harmonic oscillator of mass $m$ and classical frequency $\omega$. We obtained, for quantum numbers $r=0,1,2,3,\ldots$, 
\[
{E_r}^*  = \varepsilon _\omega  \sqrt {\frac{{\pi ^2 (r + 1)^2  - 6}}{{12}}}
\]
and
\[
{\phi_r}^* (x ) = \sqrt {\frac{1}{{L^*}(r)}} \cos \left[ {\frac{\pi }{2}\sin^2 \left(\frac{{r\pi }}{2}\right) - \frac{{(r+1)\pi x }}{2{L^*}(r)}} \right]\,,
\]
where $\varepsilon_\omega=\omega\hbar$ and
\[
\frac{1}{{{L^*}}(r)} = \sqrt{\left( {\frac{{8\alpha }}{{\pi ^2 }}\gamma (r)} \right)},\quad \gamma(r)=\sqrt { {\frac{{\pi ^2 (r + 1)^2  - 6}}{{48(r + 1)^4 }}} }\,,
\]
with $\alpha=m\omega/\hbar$.

\bigskip

The optimized eigenfunctions ${\phi_r}^* (x )$ were found to adequately describe the eigenstates of the free harmonic oscillator in the interval: \mbox{$-L^*(r)\le x\le L^*(r)$}.

\end{document}